\RequirePackage{etex}
\RequirePackage{xcolor}
\documentclass[openacc]{rstransa}%%%%where rstrans is the template name

\usepackage[shortlabels]{enumitem}
\usepackage{etoolbox}
\pretocmd{\abstractname}{\newpage}{}{}
\usepackage{mdframed}
\usepackage{algorithmic}
\usepackage{algorithm}
\usepackage{graphics}
\usepackage{graphicx}
\usepackage{multirow}
\usepackage[centerlast]{subfigure}

\usepackage{epsfig}
\usepackage{amsfonts,amsmath,amsthm}
\usepackage{threeparttable}
\usepackage{url}
\usepackage{cancel,soul,ulem}
\usepackage{changepage, calc,  enumitem, linegoal, datetime}

\newcommand{\lt}{\ensuremath <}

\usepackage{float}
\usepackage{hyperref}

\usepackage{array,booktabs}
\makeatletter
\newcolumntype{i}[1]{%
    >{\minipage[t]{\linewidth}\let\\\tabularnewline
      \itemize
      \addtolength{\rightskip}{0pt plus 50pt}% for raggedright
      \setlength{\itemsep}{-\parsep}}%
    p{#1}%
    <{\@finalstrut\@arstrutbox\enditemize\endminipage}}
\makeatother

\usepackage[pagewise]{lineno}

\begin{document}

%%%% Article title to be placed here
\title{VECMAtk: A Scalable Verification, Validation and Uncertainty Quantification Toolkit for Scientific Simulations}

\author{%%%% Author details
D. Groen$^{1,2}$, H. Arabnejad$^{1}$, V. Jancauskas$^{3}$, W. N. Edeling$^{4}$, F. Jansson$^{4,11}$, R. A. Richardson$^{2,5}$, J. Lakhlili$^{6}$, L. Veen$^{5}$, B. Bosak$^{7}$, P. Kopta$^{7}$, D. W. Wright$^{2}$, N. Monnier$^{8}$, P. Karlshoefer$^{8}$, D. Suleimenova$^{1}$, R.  Sinclair$^{2}$, M. Vassaux$^{2}$, A. Nikishova$^{9}$, M. Bieniek$^{2}$, O. O. Luk$^{6}$, M. Kulczewski$^{7}$, E. Raffin$^{8}$, D. Crommelin$^{4,10}$, O. Hoenen$^{6}$, D. P. Coster$^{6}$, T. Piontek$^{7}$ and P. V. Coveney$^{2,9}$,}

%%%%%%%%% Insert author address here
\address{$^{1}$Department of Computer Science, Brunel University London, London, UK\\
$^{2}$Centre for Computational Science, University College London, London, UK\\
$^{3}$Leibniz Supercomputing Centre, Garching, Germany\\
$^{4}$Centrum Wiskunde \& Informatica, Amsterdam, The Netherlands\\
$^{5}$Netherlands eScience Center, Amsterdam, The Netherlands\\
$^{6}$Max-Planck Institute for Plasma Physics - Garching, Munich, Germany\\
$^{7}$Pozna\'n Supercomputing and Networking Center, Pozna\'n, Poland\\
$^{8}$CEPP - Center for Excellence in Performance Programming, Atos Bull, Rennes, France\\
$^{9}$University of Amsterdam, Amsterdam, The Netherlands\\
$^{10}$Korteweg-de Vries Institute for Mathematics, Amsterdam, The Netherlands\\
$^{11}$Delft University of Technology, Delft, The Netherlands}

%%%% Subject entries to be placed here %%%%
\subject{}

%%%% Keyword entries to be placed here %%%%

%%%% Insert corresponding author and its email address}
\corres{Derek Groen\\
\email{derek.groen@brunel.ac.uk}}

\maketitle

%%%%%%%%%% Insert the texts which can accommodate on the first page in the tag "fmtext" %%%%%

\begin{fmtext}
\end{fmtext}
%%%%%%%%%%%%%%% End of first page %%%%%%%%%%%%%%%%%%%%%

%%%% Abstract text to be placed here %%%%%%%%%%%%
%\begin{abstract}
\begin{mdframed}[hidealllines=true,backgroundcolor=brown!20]
\noindent We present the VECMA toolkit (VECMAtk), a flexible software environment for single and multiscale simulations that introduces directly applicable and reusable procedures for verification, validation (V\&V), sensitivity analysis (SA) and uncertainty quantication (UQ). It enables users to verify key aspects of their applications, systematically compare and validate the simulation outputs against observational or benchmark data, and run simulations conveniently on any platform from the desktop to current multi-petascale computers. In this sequel to our paper on VECMAtk which we presented last year~\cite{groen2019introducing}, we focus on a range of functional and performance improvements that we have introduced, cover newly introduced components, and applications examples from seven different domains such as conflict modelling and environmental sciences. We also present several implemented patterns for UQ/SA and V\&V, and guide the reader through one example concerning COVID-19 modelling in detail.
\end{mdframed}
%\end{abstract}

%%%%%%%%%%%%%%%%%%%%%%%%%%%
%-------------------------------------------
% 			Introduction
%-------------------------------------------
\section{Introduction}

Computational models play an ever-growing role in predicting the behavior of real-world systems or physical phenomena~\cite{roy2011comprehensive}, using equations and/or heuristics to encode the natural laws and theories of the world we inhabit. Because models are a simplified representation of real-world systems, they can behave differently for a variety of reasons. Key model inputs, such as initial conditions, boundary conditions and important parameters controlling the model, are often not known with certainty or are inadequately described \cite{national2012assessing}. For example, in the case of human migration simulations~\cite{suleimenova2020sa}, the model may require knowledge of a wide range of input parameters, such as details of transport modes, roads, settlement populations and conflict zones, before we can execute the simulation. However, often these parameters are not precisely known or cannot be obtained with high accuracy. 

Another source of discrepancy between the model and reality are the assumptions and simplifications that are made as part of creating the conceptual model. Simplifications can reduce the computational cost of models, but also make them less accurate. And assumptions are by definition uncertain, which makes it necessary to test the model accuracy  when these assumptions are adjusted to match realistic alternative scenarios. 

The appropriate level of accuracy and reliability in the results can be obtained by ensuring \textit{not only} that the computational model accurately represents the underlying mathematical model and its solution (\textit{\emph{V}erification}) \cite{schwer2009guide,oberkampf2010verification}, but the degree to which a model is an accurate representation of the real world based on comparisons between computational results and experimental data (i.e., the deviation of the model from reality) (\textit{\emph{V}alidation}) \cite{simmermacher2015role}, and how variations in the numerical and physical parameters affect simulation outcomes (\textit{\emph{U}ncertainty \emph{Q}uantification}). Collectively, the processes involved in evaluating our level of trust in the results obtained from models are known as VVUQ. VVUQ processes provide the basis for determining our level of trust in any given model and the results obtained using it \cite{roy2011comprehensive, binois2018practical, baker2020stochastic}.

Many scientific modelling challenges involve complex and possibly multiscale, multiphysics phenomena, the results of which are unavoidably approximate. VVUQ then becomes essential to determine whether the results can be trusted, and whether decision makers can rely on them to guide subsequent actions, making our simulations \textit{actionable}. Indeed, the impact of scientific computing relies directly on its trustworthiness~\cite{oberkampf2010verification}, and VVUQ provides the basis for determining our level of trust in any given model and the results obtained using it \cite{roy2011comprehensive}.

Within this paper we describe the latest release of the VECMA toolkit. VECMA (\href{https://www.vecma.eu}{www.vecma.eu}) is an EU-funded initiative that focuses on enabling VVUQ for large-scale and complex simulations. The toolkit we present here facilitate the use of VVUQ techniques in (multiscale) applications, as well as a range of Verification and Validation Patterns (VVPs) to enable a systematic comparison of simulation results against a range of validation targets, such as benchmarks or measurements. We review a range of related research and development efforts in Section~\ref{S:related}, and then provide a brief description of the toolkit as a whole in Section~\ref{S:toolkit}. We describe the key aspects of each component in the toolkit in Section~\ref{S:components} and present a range of exemplar applications in Section~\ref{Sec:exemplars}, while we discuss on the main performance and scalability aspects of the toolkit in Section~\ref{S:scalability}. Lastly, we share our main conclusions in Section~\ref{S:conclusions}. 

\section{Related Work}\label{S:related}
Several other toolkits share a subset of the added values that VECMAtk provides. In the area of VVUQ, a well-known toolkit is Design Analysis for Optimization and Terascale Applications (DAKOTA) \footnote{https://dakota.sandia.gov} \cite{adams2009dakota}, which provides a suite of algorithms for optimization, UQ, parameter studies, and model calibration. DAKOTA is a powerful tool, but has a relatively steep learning curve due to the large number of tools available \cite{lin2012survey} and 
offers no way to coordinate resources across concurrent runs
\cite{foley2010many, elwasif2012parameter}. Similarly, there are other toolkits that help with UQ directly, such as UQTK~\cite{debusschere2018uqtk} and UQLab~\footnote{https://www.uqlab.com}. One particularly efficient method to handle UQ for a range of applications is the multi-level Monte Carlo Method~\cite{giles2015}.

In the area of VVUQ using HPC, there are several other relevant tools. OpenTURNS \cite{baudin2015open} focuses on probabilistic modelling and uncertainty management, connects to HPC facilities, and provides calibration/Bayesian methods and a full set of interface to optimization solvers. Uranie leverages the ROOT~\footnote{http://root.cern.ch} framework to support a wide range of uncertainty quantification (UQ) and sensitivity analyses (SA) activities using local and HPC resources. A key requirement for performing many types of UQ and SA is the ability to effectively run large ensembles of simulations runs. In addition to QCG-PJ, presented in this paper, there are tools such as RADICAL-Cybertools \cite{balasubramanian2019radical} that can be used to initiate and manage large simulation ensembles on peta and exascale supercomputers. In the area of surrogate modelling, GPM/SA~\cite{gattiker2008gaussian, gattiker2017gaussian} helps to create surrogate models, calibrate them to observations of the system, and give predictions of the expected system response. There is also a portfolio of available solutions for rapidly processing user-defined experiments consisting of large numbers of relatively small tasks. The examples are Swift/T\cite{wozniak2013} and Parsl\cite{babuji2019}, both of which support execution of data-driven workflows.

Another range of relevant related tools include more statistically oriented approaches. For instance, Uncertainpy \cite{tennoe2018uncertainpy} is a UQ and SA library that supports quasi-Monte Carlo (QMC) and polynomial chaos expansions (PCE) methods. PSUADE \cite{PSUADE2016} is a toolbox for UQ, SA and model calibration in non-intrusive ways \cite{hittinger2010uncertainty}, while DUE~\cite{brown2007data} assesses uncertain environmental variables, and generates realisations of uncertain data for use in uncertainty propagation analyses. PyMC3 \cite{salvatier2016probabilistic} is a Python package for Bayesian statistical modelling and probabilistic machine learning which focuses on Markov Chain Monte Carlo approaches and variational fitting. Similarly, SimLab~\footnote{https://ec.europa.eu/jrc/en/samo/simlab} offers global UQ-SA based on non-intrusive Monte Carlo methods. UQLab \cite{marelli2014uqlab} and SAFE \cite{pianosi2015matlab} are MATLAB-based tools that provide support for UQ (using e.g. PCE) and SA (using e.g. Sobol's method) respectively.

%-------------------------------------------
% 			The VECMA toolkit
%-------------------------------------------
\section{The VECMA toolkit (VECMAtk)}\label{S:toolkit}
The main objectives of VECMAtk are to facilitate the implementation of (a) uncertainty quantification and sensitivity analysis patterns (UQPs) and (b) verification and validation patterns (VVPs). The UQP implementations automate routines for uncertainty quantification and sensitivity analysis, while the VVP implementations enable verification and validation procedures for high performance (multiscale) computing applications. Ye et al. present the concepts of UQP on an algorithmic level~\cite{Ye2020UQP}, while an algorithmic paper on VVPs is still in preparation. Both implementations support remote execution on petascale and emerging exascale resources, as well as execution on local machines. In addition, we support a range of existing VVUQ mechanisms and provide tools for users to facilitate the adoption of VVUQ in their applications. As we want our toolkit to be general-purpose, taken up by users, and  flexible, we have four main factors that shape our development approach:

\begin{enumerate}
	\setlength\itemsep{-0.1em}
	\item the need to fit into existing applications with minimal modification effort,
	\item the need to support any application domain,
	\item the flexible and recombinable nature of the toolkit itself,
	\item the geographically distributed nature of the users and particularly the developers.
\end{enumerate}

As a result, we position VECMAtk as a collection of elements that can be reused and recombined in different workflows, interlinked with stable interfaces, data formats and APIs, to facilitate VVUQ in any application. VECMAtk specifically allows users with (multiscale) applications to combine generic, lightweight patterns to create a system-specific VVUQ approach that covers all relevant space and time scales, including the scale-bridges between them. The exemplar applications (which we present in Section~\ref{Sec:exemplars}) map to all three of the established multiscale computing patterns: Extreme Scaling (ES), Replica Computing (RC) and Heterogeneous Multiscale Computing (HMC)~\cite{alowayyed2017multiscale}.

%-------------------------------------------
%  Changes compared to the initial release
%-------------------------------------------
\subsection{Release schedule and changes compared to the initial release}

The VECMA toolkit is an open-source and open development project, meaning that the latest source and development notes can be accessed at any time. In terms of releases, we have adopted a schedule with minor and major releases. Minor releases are made publicly every 3 months, are advertised within the project, and have limited amount of additional documentation and examples. The first major release was made in June 2019; the ensuing two releases will be made in June 2020 and June 2021, and will be public and fully advertised. They are accompanied with extensive documentation, tutorials, examples, training events and dedicated uptake activities. In addition to these two release types, we provide informal periodic updates to the master code branches, documentation and the integration of the components.

Since our first major release in June 2019, all VECMAtk components have improved scalability, a range of new features and capabilities, extended and revised application tutorials, improved technical documentation, and a wide range of user-requested bugfixes. We present the fundamental improvements and changes in the development of each VECMAtk component since the first major release as part of Appendix A.

%-------------------------------------------
% 			Overview of the VECMAtk components
%-------------------------------------------
\section{Overview of the VECMAtk components}\label{S:components}
\begin{figure}[ht]
	\centering
		 % 								trim = left bottom right up	 	
		\includegraphics[scale=0.6]{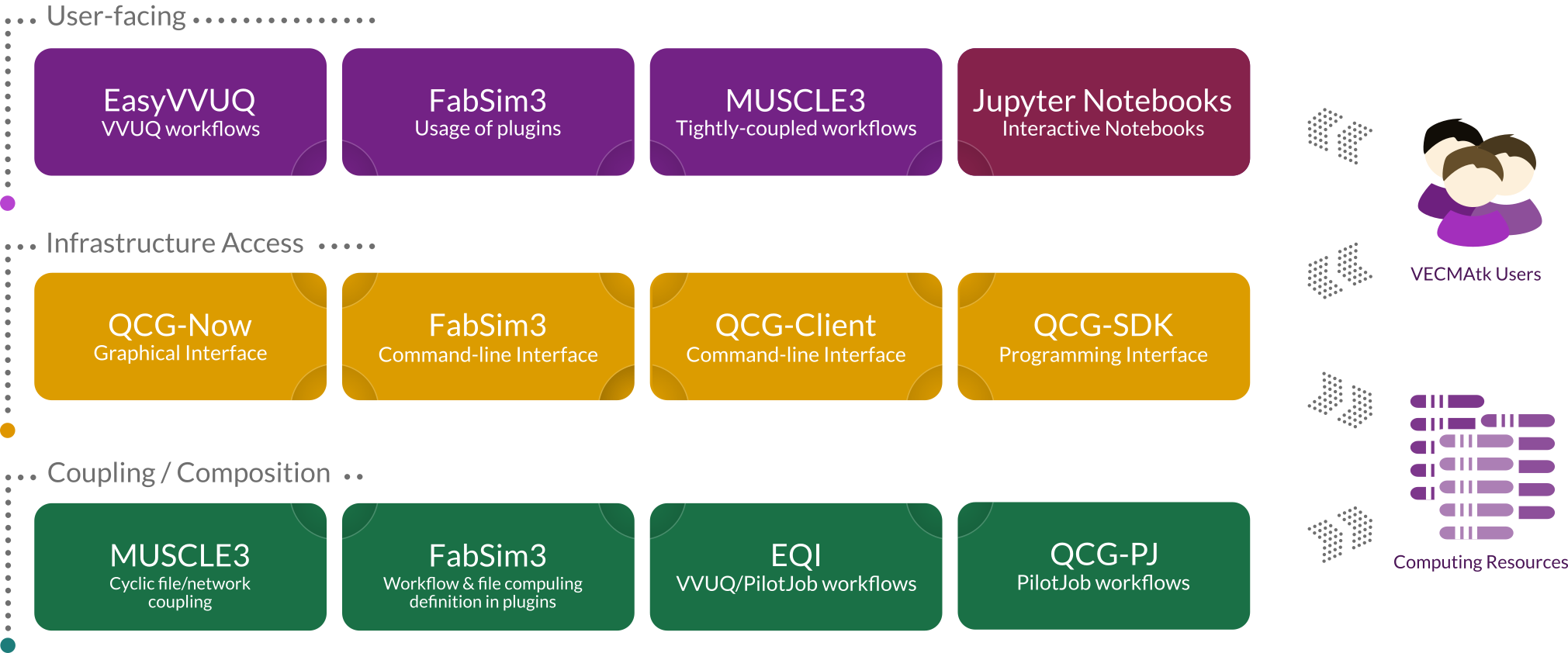}
	\caption{Overview of the main components, and their role within the VECMA toolkit.}	
	\label{Fig:comp}
	\medskip % induce some separation between caption and explanatory material
	\end{figure}

We present a graphical overview of the various components in VECMAtk in Fig.~\ref{Fig:comp}. VECMAtk contains four main components: \textit{EasyVVUQ}~\cite{easyvvuq2020} simplifies the implementation and use of VVUQ workflows with focus on large scale scenarios, \textit{FabSim3}~\cite{groen2016fabsim} helps to automate computational research activities,  \textit{MUSCLE 3}~\cite{MUSCLE3} supports the coupling multiscale applications, and the \textit{QCG tools}~\footnote{https://www.qoscosgrid.org} facilitate advanced workflows, and the process of design and execution of applications, using HPC infrastructures. All these components can be flexibly combined with other VECMAtk and third party components to create diverse application workflows. In addition, no single component is essential to all applications: for example FabSim3 and QCG-Client are to a limited degree interchangeable, and at least one application fully relies on a third party tool (the Mogp emulator~\footnote{https://github.com/alan-turing-institute/mogp\_emulator}) instead of EasyVVUQ.

In this section we summarize each of the components that comprise VECMAtk, while we refer to the VECMAtk website \href{https://www.vecma-toolkit.eu/}{https://www.vecma-toolkit.eu/} for more detailed technical information.

%-------------------------------------------
% 			EasyVVUQ
%-------------------------------------------
\subsection{EasyVVUQ}
%\desc{The input will be provided by Vytautas, and Jalal}

EasyVVUQ~\cite{Richardson2020easyvvuq}\cite{easyvvuq2020} is a Python library designed to facilitate implementation of advanced non-intrusive VVUQ techniques, with a particular focus on high performance computing, middleware agnosticism, along with single scale and multiscale modelling. 
%EasyVVUQ is focused on modularity, scalability and ease of use. 
Generally, non-intrusive VVUQ workflows involve sampling the parameter space of the simulation. This usually means running the simulation multiple times. This process is tedious and error prone if done in an \textit{ad hoc} manner, and further exacerbated if the parameter space is large with a corresponding large number of runs (e.g. of order thousands to millions of runs), or if each simulation is very computationally expensive. 
EasyVVUQ allows one to coordinate the whole process - from sample generation through execution to analysis - although the execution stage is external to EasyVVUQ, and can be done  by tools such as QCG-PilotJob, FabSim3 or Dask, or even manually. EasyVVUQ seeks to be agnostic to the choice of execution middleware, due to the large range of possible execution patterns that may be required in an HPC workflow. Additionally, the library carefully tracks and logs applications of the sampling elements along the way, allowing a reasonable level of restartability and failure resistance.

EasyVVUQ breaks down the VVUQ process into five distinct stages.

\begin{enumerate}
    \item Application description, which further can be divided into the following items:
    \begin{enumerate}
        \item Encoder - responsible for producing input files for the simulation. 
        \item Decoder - responsible for parsing the output files of the simulation and extracting the needed values. 
        \item Execution action - describes how the simulation is run. 
    \end{enumerate}
    \item Sampling - this step is very dependent on the VVUQ technique in question. The main task of sampling components is to produce a list of parameter sets for the simulation.
    \item Execution - this is handled entirely externally to EasyVVUQ. 
    \item Collation - collects the outputs of the simulations and then stores them in the EasyVVUQ database, to be retrieved later for analysis.
    \item Analysis - perform statistical analysis on the collected data. This analysis can then inform further actions such as collecting more samples. 
\end{enumerate}

EasyVVUQ currently has support for variance-driven sensitivity analysis, stochastic collocation and polynomial chaos expansion methods, as well as more basic statistical analysis methods such as bootstrapping. Additionally, all components of EasyVVUQ are easily extensible, allowing users to customize every stage of the VVUQ workflow to suit their needs. Nevertheless, for many applications the built-in components provide out-of-the-box functionality, requiring the user only to define input and output formats and perform analysis. 

%-------------------------------------------
% 			FabSim3
%-------------------------------------------
\subsection{FabSim3}

FabSim3\footnote{https://github.com/djgroen/FabSim3}~\cite{groen2016fabsim} is a Python-based toolkit for automating complex computational research activities which has several aims: 
First it helps to automate and simplify the creation, management, execution and modification of complex application workflows. To do so, FabSim3 supports functionalities such as ensemble runs, remote executions, iterative processing and code couplings. Second, it aims to help researchers become quicker and more systematic in handling complex applications in particular. To support this, FabSim3 is designed to be easy to customize for use on new machines and with new simulation codes, and automatically curates a wide range of environment and state variables to aid the testing and debugging of application runs. As usage examples, researchers may want to run and rerun static configurations, run a range of slightly different workflows, define new types of complex applications altogether or construct a routine to automatically validate their code.

\begin{figure}[ht]
	\centering
		 % 								trim = left bottom right up	 	
		\includegraphics[trim=10 10 10 10,clip=true,scale=0.40]{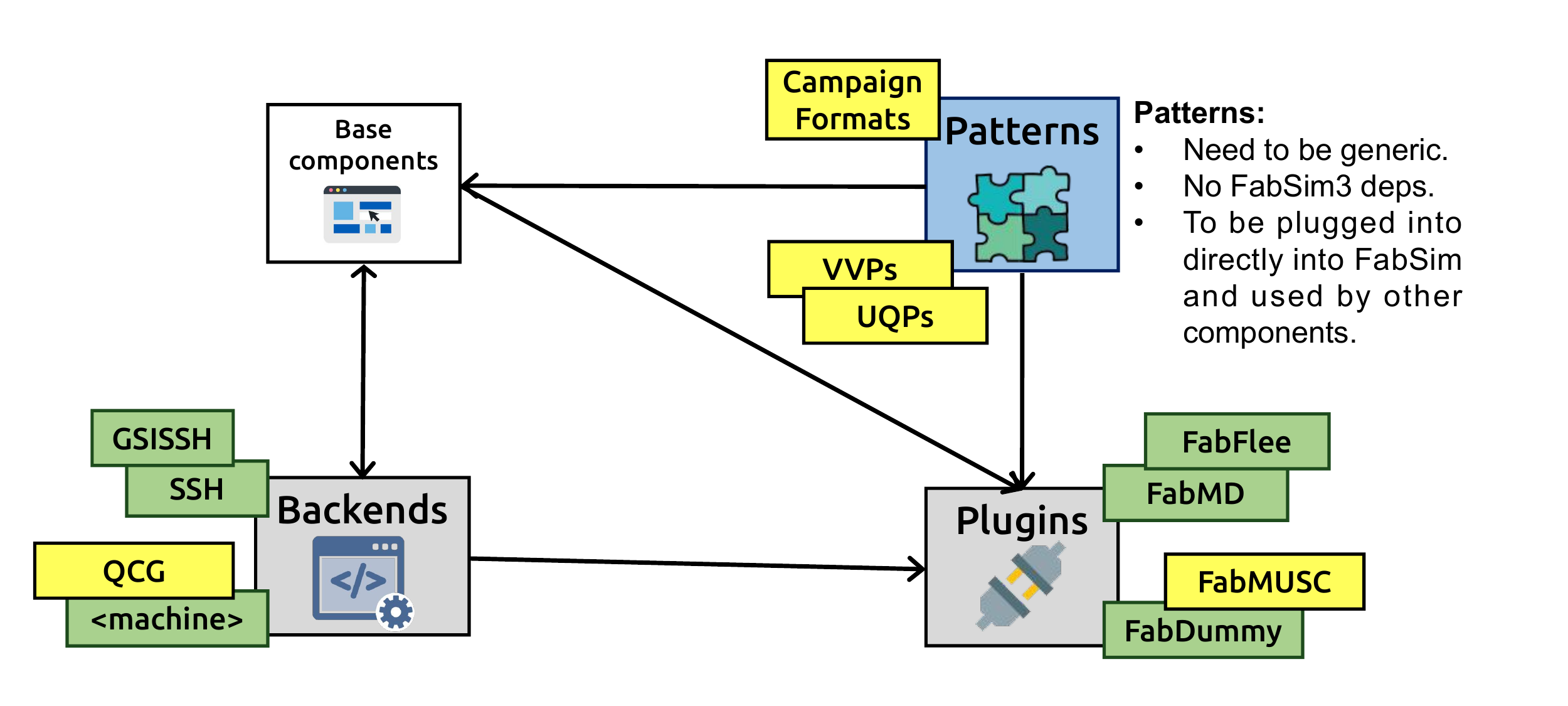}
	%\caption{VECMAtk : System component and semantic workflows.}
	\caption{High-level overview of the FabSim3 architecture, showcasing a few of the key plugins, patterns and back-end functionalities available in the tool. Boxes in green are completed, while boxes in yellow are working, but subject to further extensions and improvements.}	
	\label{FabSim3_components}
	\medskip % induce some separation between caption and explanatory material

	\end{figure}

FabSim3 also supports a range of adaptable mechanisms for code coupling, and domain-agnostic code patterns that provide building blocks for enabling VVUQ in new applications.  We provide an overview of the FabSim3 architecture in Figure~\ref{FabSim3_components}.

In the context of VECMAtk, FabSim3 plays a key role in introducing application-specific information in the Execution Layer, enabling users to combine different UQPs and VVPs, and providing an approach to curate large sets of production runs. Also, FabSim3 can use QCG-Client or QCG-SDK to submit ensemble runs via a "pilot job" mechanism~\cite{groen2019introducing,luckow2012towards} (see Subsection~\ref{sec:QCG-PJ}), which boosts efficiency by starting and managing sub jobs without each of them needing to individually wait for resources to become available.

%-------------------------------------------
% 			MUSCLE 3
%-------------------------------------------
\subsection{MUSCLE 3}\label{sec:MUSCLE3}
%\desc{The input will be provided by Lourens}

The third version of the Multiscale Coupling Library and Environment\cite{MUSCLE3}\cite{MUSCLE3iccs2020} (MUSCLE 3) is intended to simplify the coupling of temporally or spatially scale-separated submodels into a single (distributed) simulation. At run-time, the submodels are started in parallel, and exchange information via the network. Submodels and other simulation components are unaware of each other as MUSCLE 3 delivers the messages to receivers given in a configuration file. The Multiscale Modelling and Simulation Framework (MMSF, \cite{BORGDORFF2013465,Chopard2014}) may be used to determine how the submodels should be coupled given their relative spatial and temporal scales. %MUSCLE 3 is capable of coupling scale-overlapping multiphysics simulations.
%but such tight coupling is probably better left to frameworks specifically designed for this.

MUSCLE 3 supports time-scale separated macro-micro couplings, including automatically re-running the micro model as often as needed to match the macro model's time steps. It also supports space scale separation, which entails coupling a single macro model to a set of micro models, via support of sets of multiple submodel instances and special vector ports that can be used to communicate with them. Model settings (parameters and technical configuration) are put in the single configuration file, and are transmitted to the individual instances automatically. Components may be inserted into the simulation that generate a parameter overlay, which combined with vector ports and sets of instances yields an ensemble for e.g. UQ. MUSCLE 3 supports a range of UQPs, including those for semi-intrusive UQ \cite{NIKISHOVA201980,Ye2020UQP}.   
With MUSCLE 3, this can be done by changing the configuration file to add a few more components and modifying the connections; the submodels can remain unaltered.

\subsection{QCG Tools}
The QCG suite of tools help to facilitate advanced workflows, and the process of design and execution of applications, using HPC infrastructures. Four specific QCG components form part of VECMAtk: QCG-Client, QCG-PJ, EQI and QCG-Now, which we now describe in turn.

%-------------------------------------------
% 			QCG-Client
%-------------------------------------------
\subsubsection{QCG-Client}
QCG-Client is a command line tool to access remote computing infrastructure. The tool allows submission of different types of jobs, including complex workflows, parameter sweep tasks and array jobs, on single or multiple clusters. It uses the QCG-Broker service to manage the execution of workflows, e.g. through multi-criteria selection of resources. QCG-Client can be deployed as a container and is in this form integrated with FabSim3. This allows users to select from both command-line tools, QCG-Client and FabSim3, when they access QCG services.

%-------------------------------------------
% 			QCG Pilot Job
%-------------------------------------------
\subsubsection{QCG Pilot Job (QCG-PJ)}\label{sec:QCG-PJ}
To perform UQ procedures, we must be able to flexibly and efficiently execute a large number of simulations. This is because we need a large and dynamically created parameter-space, which in turn feeds into an ensemble of model executions to get statistically correct results. Within VECMAtk, we use QCG-PilotJob (QCG-PJ) by default to fulfil this requirement, primarily because it can be installed automatically by users without admin rights, and because of its scalability and limited dependency footprint. For instance, QCG-PJ can be set up by users without requiring other components from the QCG environment.

QCG-PJ is designed to schedule and execute many small jobs inside one scheduling system allocation on an HPC resource. Direct submission of a large group of jobs to a scheduling system can result in a long completion time as each single job is scheduled independently and waits in a queue. In addition, the submission of a group of jobs is often restricted (and sometimes even prohibited) by system administrators. Some systems do support bespoke job array mechanisms, but these mechanisms normally only allow jobs with identical resource requirements. Jobs that are part of a multiscale simulation or application workflow by nature vary in requirements and therefore need more flexible solutions. 

To use QCG-PJ a user submits a job with a QCG-PJ Manager instance, which is executed like a regular job. However, unlike regular jobs this job internally manages a user-defined workflow consisting of many sub-jobs. The manager executes commands to submit jobs, cancel jobs and report resources usage; it may either be preset or listen dynamically to requests from the user. QCG-PJ then manages the resources and jobs in the system, taking into account resource availability and mutual dependencies between jobs. Users can interact with QCG-PJ either using a file-based or network-based interface. The file-based approach works well for static scenarios where the number of jobs is known in advance, while the network interface allows users to dynamically send new requests and track execution of previously submitted jobs at run-time. Although intended for HPC resources, QCG-PJ Manager also supports a local execution mode to allow users to test their scenarios on their own machines (with requiring a scheduling system allocation). 

%\begin{figure}[ht]
%	\centering
		 % 								trim = left bottom right up	 	
%		\includegraphics[scale=0.70]{./Figs/qcgpj.png}
%	\caption{User perspective sketch of QCG-PJ. The blue boxes are the user-defined tasks run in a single allocation. QCG-PJ Manager manages this allocation, and removes the need for users to monitor and interact with these tasks directly.}	
%	\label{qcgpj}
%	\medskip % induce some separation between caption and explanatory material

%	\end{figure}

%-------------------------------------------
% 			EasyVVUQ-QCGPilotJob
%-------------------------------------------
\subsubsection{EQI}

Many workflows involving EasyVVUQ require a large number of jobs to be executed, and EasyVVUQ delegates this responsibility to other tools (either existing, or user created).  QCG-PJ is the main tool available in VECMAtk to fulfil that purpose, and it offers a flexible but generic API that is less appropriate for specific use cases. To provide an API that is specifically adjusted to EasyVVUQ, we have developed EQI, which is an acronym for the (E)asyVVUQ-(Q)CG-PJ (I)ntegration API. This API introduces domain-oriented concepts, such as predefined tasks for execution and encoding, that facilitate the easy integration of EasyVVUQ workflows with QCG-PJ. With the API, users can perform VVUQ computations with QCG-PJ, invoking HPC resources. The development life-cycle of EQI is synchronized with the releases of both EasyVVUQ and QCG-PJ, to ensure a persistently up-to-date API.

%-------------------------------------------
% 			QCG-Now
%-------------------------------------------
\subsubsection{QCG-Now}

\begin{figure}[ht]
	\centering
		 % 								trim = left bottom right up	 	
		\includegraphics[scale=0.18]{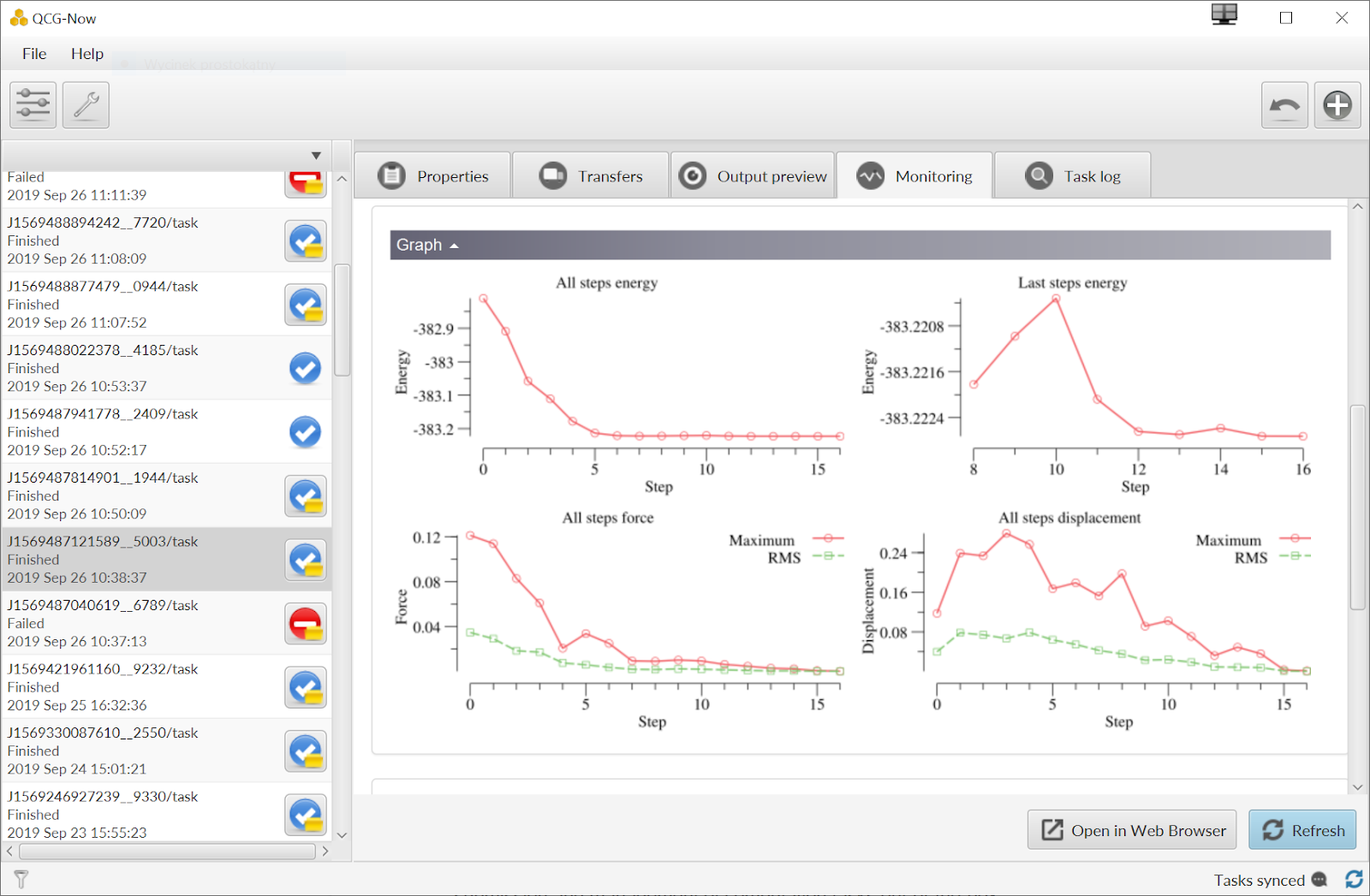}
	\caption{QCG-Now with the embedded QCG-Monitoring view. The integration with QCG-Monitoring allows users to track progress of execution of applications in a graphical way, for example presenting dynamically updated tables, images or, as in this screenshot, charts.}	
	\label{QCG_now}
	\medskip % induce some separation between caption and explanatory material
	\end{figure}

Since the traditional command-line interfaces may be perceived by many users to be too complicated, within VECMAtk we decided to provision a easy-to-use graphical tool for users. QCG-Now is graphical desktop program that enables the submission and management of jobs on HPC resources. It also supports automatic notifications, data sending and receiving. Since the first release of QCG-Now last year, several improvements have been made: for instance it is now integrated with the QCG-Monitoring system which allows users to track progress of execution of their tasks directly from its graphical interface. This capability is particularly important for VECMAtk users who want to keep track of the progress of long-lasting executions of EasyVVUQ and QCG-PJ workflows. Another new feature allows users to store frequently used execution schemes as quick templates for easy reuse. We present a graphical demonstration of the tool in Fig.~\ref{QCG_now}.

%-------------------------------------------
% 	VECMAtk workflows
%-------------------------------------------
\subsection{VECMAtk workflows}	
\begin{figure}[ht]
	\centering
		 % 								trim = left bottom right up	 	
		\includegraphics[scale=0.65]{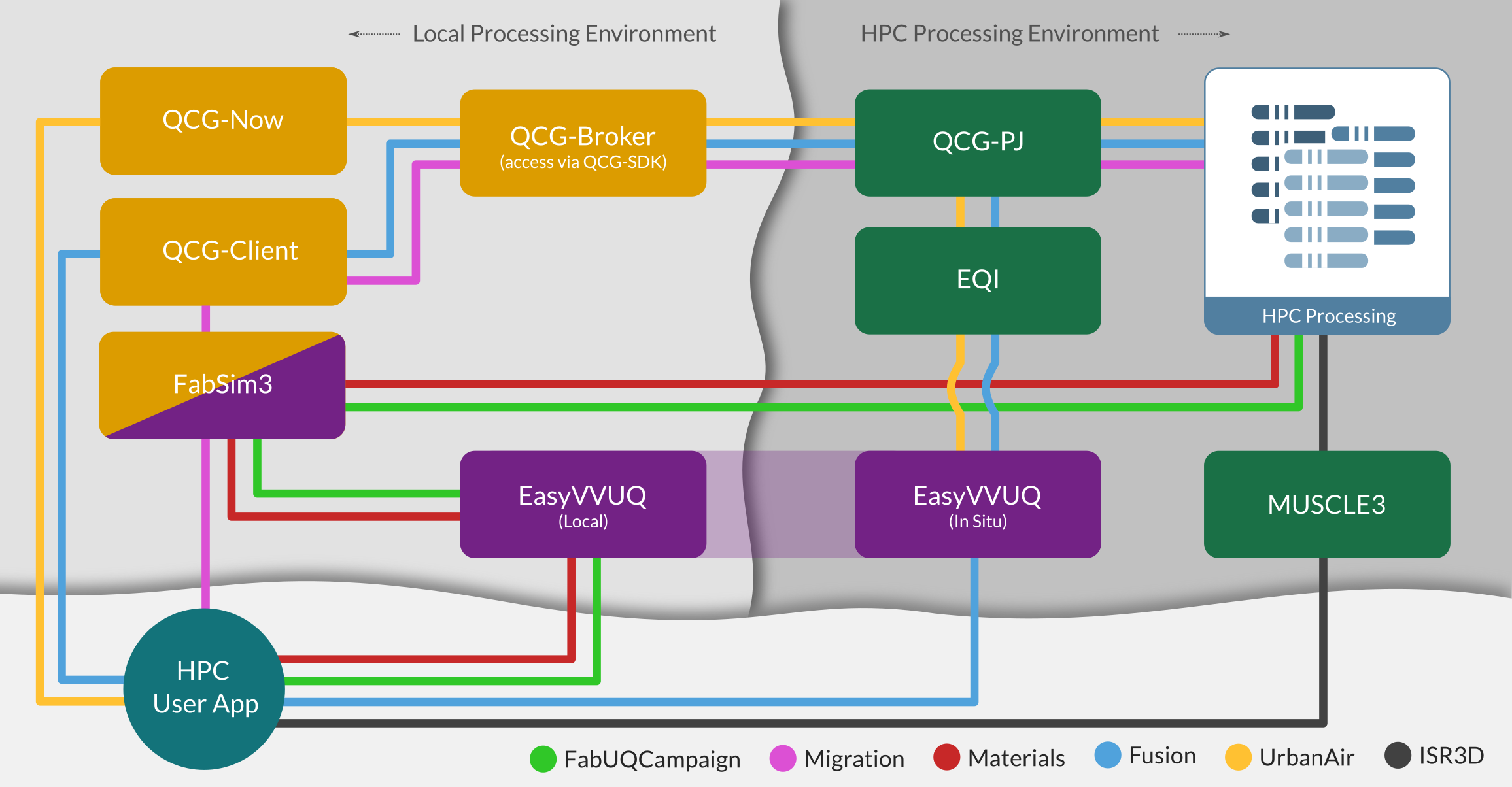}
	%\caption{VECMAtk : System component and semantic workflows.}
	\caption{“Tube map” showing which VECMAtk components are used in the various exemplar applications. VECMAtk components are given in boxes, and the application tutorials are indicated using coloured lines. Note that code using the EasyVVUQ~\cite{Richardson2020easyvvuq} library may be located either on the local desktop for ease of use or on a remote HPC resource for improved performance. Source: \href{https://www.vecma-toolkit.eu/toolkit/}{https://www.vecma-toolkit.eu/toolkit/} }	
	\label{VECMAtk_tube_map}
	%\medskip % induce some separation between caption and explanatory material

	\end{figure}

In Figure \ref{VECMAtk_tube_map} we show the main components and a few examples as to how they have been combined, leveraging added values where relevant whilst maintaining a limited deployment footprint. These components can be combined, but also integrated with third party components.

Up to now, the following combinations have been particularly common:
\begin{itemize}
	\setlength\itemsep{-0.1em}
	\item FabSim3 + QCG-Client : enables users to submit their job to QCG-Broker to schedule them across multiple remote (QCG-supporting) machines. Additionally, FabSim3 tool support other job scheduling system such as SLURM \footnote{https://slurm.schedmd.com/overview.html}, Portable Batch System (PBS) \footnote{https://www.pbspro.org/}.

	\item FabSim3 + EasyVVUQ : enables users to automate the VVUQ processes into their applications workflow. After creating their EasyVVUQ campaigns, users can convert them to FabSim3 ensembles using a one-liner (\texttt{campaign2ensemble}) command and submit the ensemble jobs through FabSim3. Then they convert results back to EasyVVUQ using the \texttt{ensemble2campaign} command, where it is decoded and further analyzed.

	\item FabSim3 + QCG Pilot Job : enables users to create and manage pilot jobs using FabSim3 automation.

	\item EasyVVUQ + QCG Pilot Job + EQI: enables EasyVVUQ users to execute their tasks directly using pilot jobs.
\end{itemize}

There are also several examples of integrations with third party components. For instance, there is an application example that uses the mogp emulator in place of EasyVVUQ~\footnote{https://github.com/alan-turing-institute/mogp\_emulator}, and a working EasyVVUQ example that relies on Dask instead of QCG-PilotJob~\footnote{https://easyvvuq.readthedocs.io/en/dev/dask\_tutorial.html} for ensemble job submission. Other integration are indeed possible as well, and the inclusion of a different job management and execution engine may be beneficial for individual applications and/or resources. To name a few examples, RADICAL-Cybertools can offer exceptional performance on machines where it is set up~\cite{balasubramanian2019radical}, Swift/T\cite{wozniak2013} can facilitate complex data flows with minimal serialization while Parsl\cite{babuji2019} is optimised for the fast and flexible execution of Python programs and Jupyter notebooks.

As indicated in Fig.~\ref{Fig:comp} earlier, there are four components that provide a suitable entry point for new users to the toolkit. EasyVVUQ facilitates users that initially focus on incorporating VVUQ in their simulations. FabSim3 is aimed at users that initially focus on enabling complex and curated workflows, which could involve ensemble or dynamic execution. QCG-Now is intended for users that prefer a graphical interface for managing their simulations and VVUQ workflows. And MUSCLE3 is well-suited for users whose primary concern is multiscale code coupling.

%-------------------------------------------
% 			Exemplar applications
%-------------------------------------------
\section{Exemplar applications}\label{Sec:exemplars}	
With VECMAtk we aim to provide a toolkit that can be applicable to any domain/application where UQ and SA are required, and computational modelling and simulation can be applied. To achieve this aim, we rely on an increasing number of leading-edge scientific applications to support the testing and development of the toolkit. Within this paper, we present a selection of these applications which cover the following topics: (a) future energy sources, (b) human migration, (c) climate, (d) advanced materials, (e) urban air pollution and (f) biomedicine and (g) epidemiology. For each application we show how the toolkit provides added value in terms of enabling UQ. In addition, we describe how we use Verification and Validation Patterns (VVPs) to help facilitate V\&V for the first two applications. 
In each of these concise application descriptions, we focus on how VECMAtk is being used by a range of different applications, which components are primarily employed/invoked, and what primary benefits the toolkit delivers for the application users.

%-------------------------------------------
% 			Fusion
%-------------------------------------------
\subsection{Fusion}

Thermonuclear fusion has the potential of becoming a new source of energy that is carbon-free.  The dynamics of fusion plasmas span a wide range of scales in both time and space, as for example micro-instabilities formed in plasma turbulence can interrupt the overall macroscopic transport and destroy confinement.  To simulate such phenomena, we have built a multiscale fusion workflow that includes equilibrium, turbulence and transport physics to model the plasma dynamics \cite{luk2019compat}.

Our central purpose is to ensure that the results are reproducible and can become a reliable representation of the experiment measurements. To achieve this, it is key to use VVUQ approaches.

\begin{figure}[ht]
    \centering
    \includegraphics[scale=0.60]
    {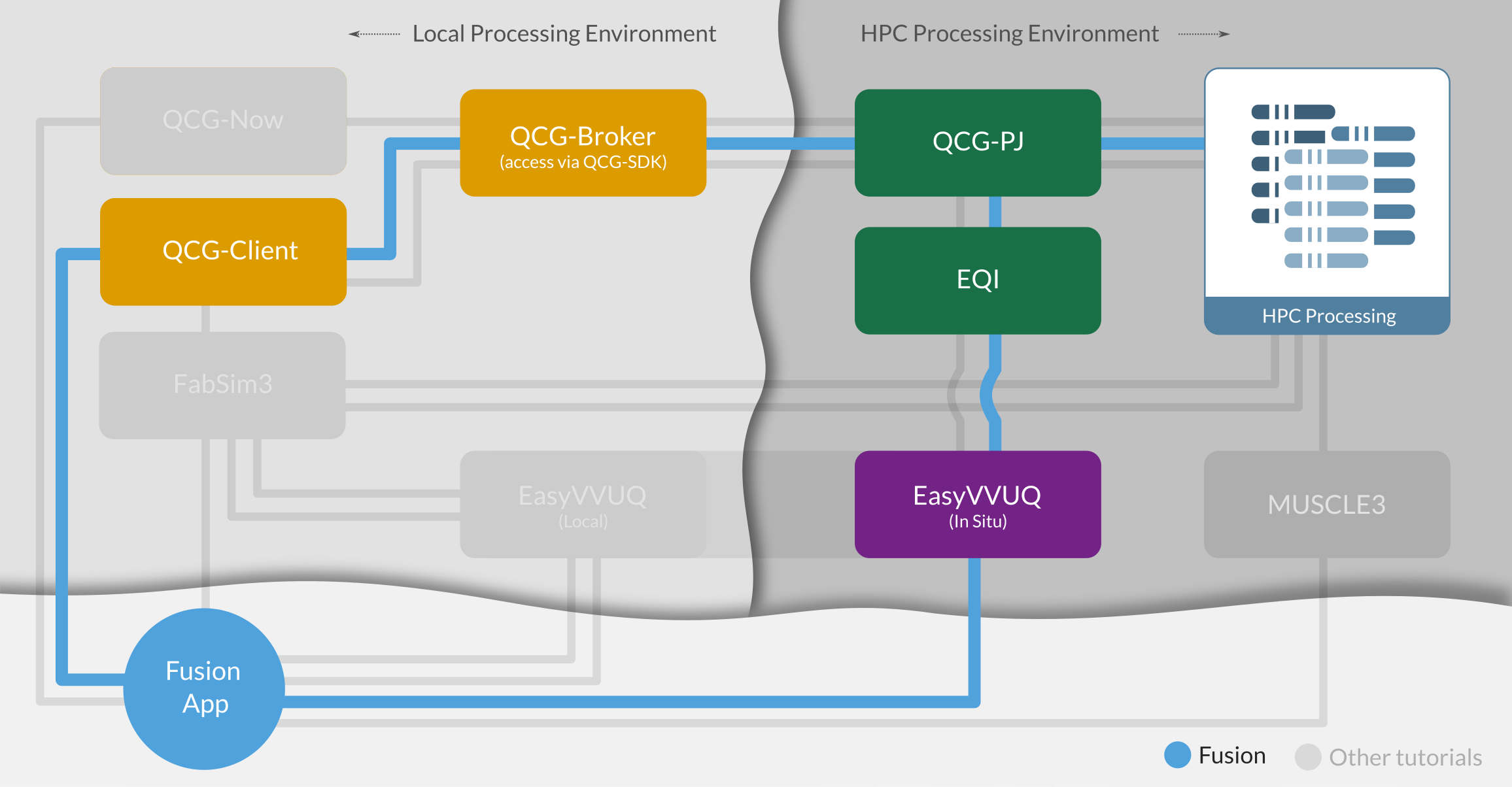}
	\caption{“Tube map” showing which VECMAtk components are used by the Fusion application.}	
	\label{FusionMap}
	\medskip
\end{figure}

We integrated UQ methods using the EasyVVUQ library and  performed a  parallel  execution  of  samples  in  a single batch allocation with QCG-PJ through the EQI. Those tools, available as part of the VECMA toolkit (see Figure \ref{FusionMap}) and  in  conjunction  with  the  coupled  workflow  nature  of  the fusion application,  have  proven  to  be  simple  to  use  and  to  allow  a  large  variety  of experiments with different methods and models (please see \cite{Lakhlili2020} for further details). We also performed a validation using the~\texttt{ValidationSimilarity} VVP, which is implemented in EasyVVUQ and help us find similarities between the probability distributions of the quantities of interest in our simulation results and in the experimental measurements \cite{Luk2020}. For this purpose we can choose between several relevant metrics, such as  Hellinger distance \cite{nikulin2001hellinger}, Jensen–Shannon distance which is a symmetrized and smoothed version of the Kullback–Leibler divergence \cite{kullback1951} and Wasserstein metrics \cite{Villani16}.

%-------------------------------------------
% 			Forced human migration
%-------------------------------------------
\subsection{Forced human migration}
Forced migration reached record levels in 2019, and forecasting it is challenging as many forced population data sets are small and incomplete, and armed conflicts are often unpredictable in nature \cite{groen2016refugee}. Nevertheless, forced migration predictions are essential to improve the allocation of humanitarian support by governments and NGOs, to investigate the effects of policy decisions and to help complete incomplete data collections on forced population movements. Through the use of the FLEE \footnote{https://github.com/djgroen/flee-release} agent-based simulation code, we model and forecast the distribution of incoming refugees across destination camps for African countries \cite{suleimenova2017prediction}. 

The VECMA toolkit provides us with the ability to automatically construct, execute and analyse ensemble simulations of refugee movements \cite{suleimenova2017automation}, to efficiently compute the sensitivities of key parameters in our simulation \cite{suleimenova2020sa}, to couple different types of models to form multiscale workflows~\cite{groen2019towards} and to facilitate the rapid execution of large job ensembles on supercomputers. We also use a VVP named \texttt{EnsembleValidation} to systematically validate our simulations against data from a range of conflicts (see e.g. here~\cite{suleimenova2020policy}). This VVP performs a validation on the output directory of each run and uses an aggregation function to combine all output validation results into a combined metric.

\begin{figure}[ht]
	\centering
		 % 								trim = left bottom right up	 	
		\includegraphics[scale=0.6]{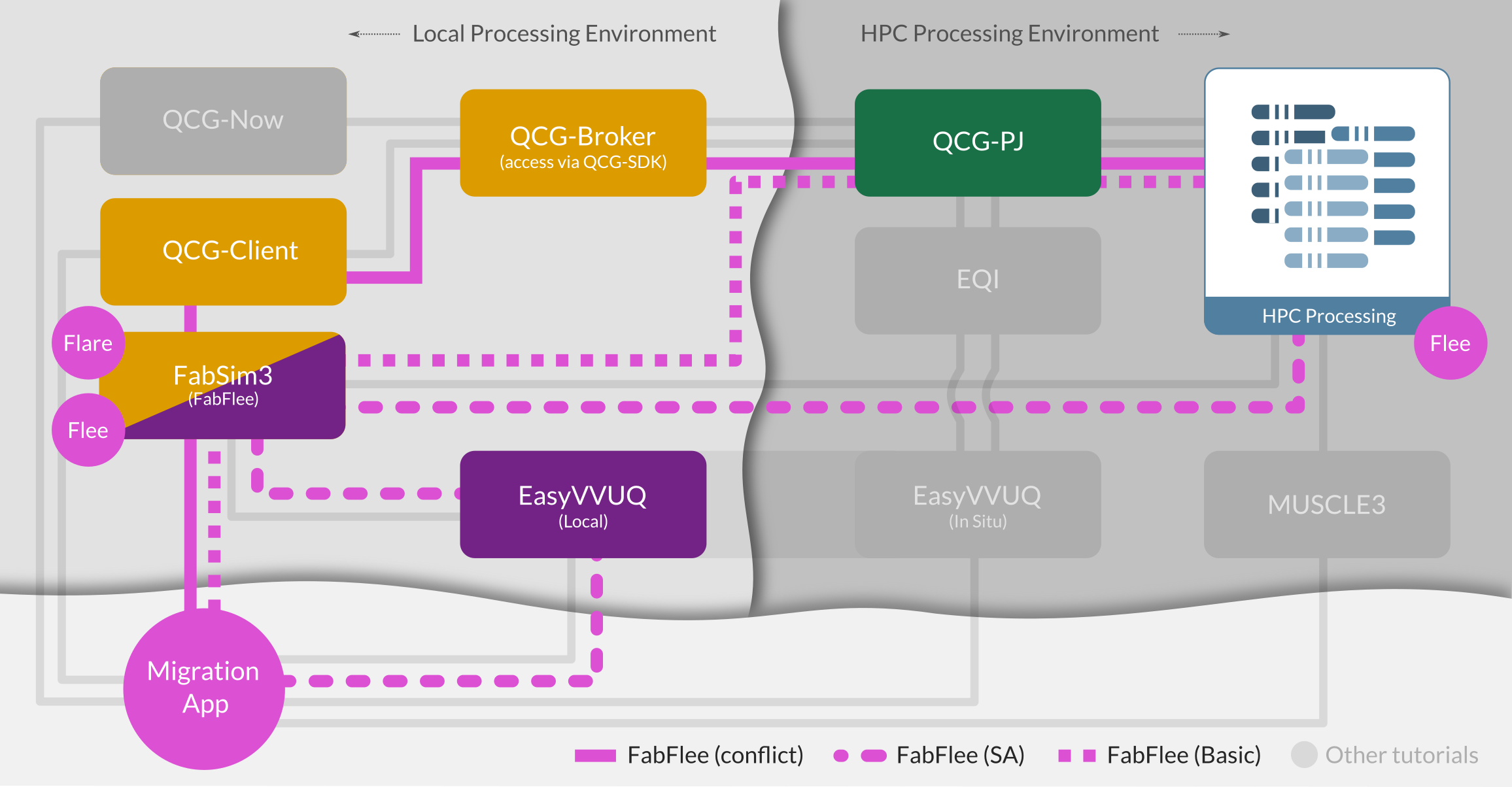}
	%\caption{Overview of the FabFlee plugin path in congestion with other components.}	
	\caption{Tube Map showing which VECMAtk components are used in the FabFlee plugin. VECMAtk components are given in boxes which enable users to include their added values while retaining a limited deployment footprint. The black and grey lines define the FabFlee path using VECMAtk components for execution.}

	\label{FabFleeMap}
	\medskip % induce some separation between caption and explanatory material
	\end{figure}

We summarize our main application workflows in Figure~\ref{FabFleeMap}. We rely on the FabFlee plugin in FabSim3 to automate the simulation activities required to analyse different policy decisions, to provide the VVP functionality and to facilitate coupling with other models and data sources. Examples include an acyclic coupling to a conflict model~\cite{groen2019towards}, but we are also working on a cyclic macro-micro model for the South Sudan conflict. We also use EasyVVUQ with FabFlee to perform sensitivity analyses for varying agent awareness levels, speed limits of refugee movements, location move chances and other simulation parameters~\cite{suleimenova2020sa}. Lastly, we use QCG Pilot Job to facilitate the very large number of runs required to do this analysis, which is required for our more advanced workflows that so far involved up to 16,384 jobs.

%-------------------------------------------
% 			Climate
%-------------------------------------------
\subsection{Climate}

For climate modelling, the range of space and time scales present in (geophysical) turbulent flow problems poses challenges, as this range is too large to be fully resolved in a numerical simulation. As such, micro scale effects on the resolved macroscopic scales must be taken into account by empirical parameterizations, see e.g.\ \cite{gent1990isopycnal}. These parameterizations often involve a number of coefficients, for which the value is only known in an approximate manner. In addition, they are subject to so-called model error or model-form uncertainty, which is the uncertainty introduced due to the assumptions made in the mathematical form of the parameterization scheme.

\begin{figure}
    \centering
    \includegraphics[scale=0.60]{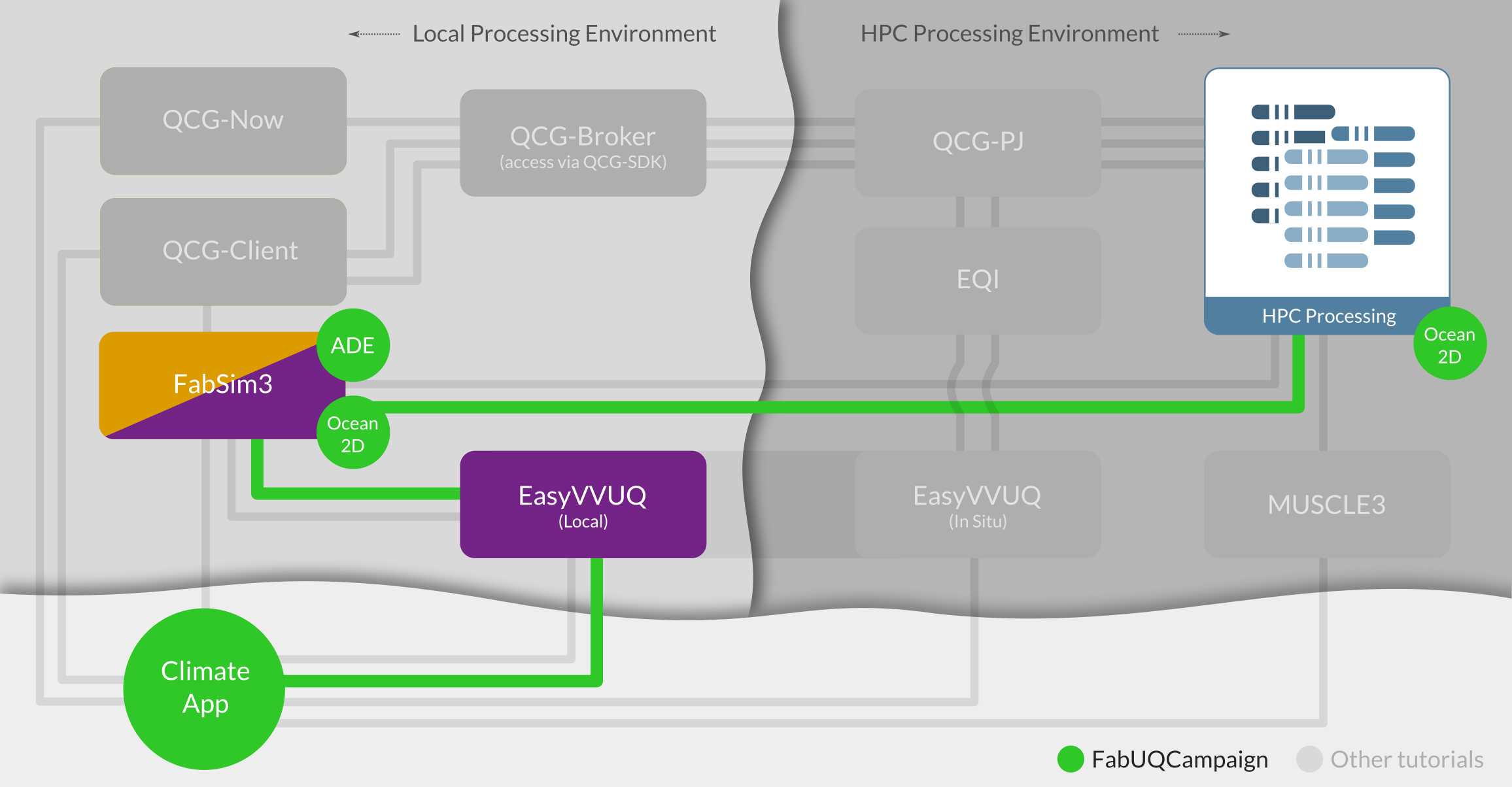}
    \caption{A Tube Map showing the VECMA components used in the climate application. Here "FabUQCampaign" is a plugin used to run an ensemble of EasyVVUQ samples on HPC resources. Furthermore, "ADE" and "Ocean2D" are two example applications (advection diffusion equation and a two-dimensional ocean model) for which tutorials can be found in \cite{edeling2019fabuqcampaign}.}
    \label{fig:FabUQMap}
\end{figure}

Within VECMA we focus on the uncertainty in time-averaged climate statistics of geophysical flow models, due to various assumptions made in the parameterizations. EasyVVUQ allows us to asses the uncertainty in the output statistics due to the imperfectly known coefficients. In addition, we currently use EasyVVUQ for uncertainty quantification of an established local atmospheric model, DALES \cite{Heus2010, Jansson2020}, which is used to simulate atmospheric processes including turbulence, convection, clouds, and rain. Here, we quantify the uncertainty in model output from different sources, including uncertain physical input parameters, model choices and numerical settings, and the stochastic nature of turbulence modelling. To run the EasyVVUQ ensemble of simulations on HPC resources we use FabSim3, see Figure \ref{fig:FabUQMap}. To address the more fundamental model-form uncertainty we are currently investigating the use of data-driven stochastic surrogates to replace traditional deterministic parameterizations; see e.g.\ \cite{edeling2020reducing, crommelin2020resampling} for recent results.

%-------------------------------------------
% 			Materials
%-------------------------------------------
\subsection{Advanced materials}

Developing novel advanced materials can be a very expensive and time consuming process, often taking over 20 years from initial discovery to application \footnote{https://www.mgi.gov/}. Chemical and material modelling techniques are well developed for single scales, but engineering applications require understanding across many scales~\cite{vassaux2020toward, vassaux2019heterogeneous, suter2020}. Making actionable predictions with these modelling techniques, to improve on the laborious experimental development process, will require harnessing multiple simulation techniques and providing tight error bars on their predictions. 

Using VECMAtk, one can generate, execute, collate, and analyse ensemble simulations of mechanical tests in an automatic fashion. This allows us to effectively gather statistics on a system of interest and explore an input parameter space with minimal human oversight.

To do this we use EasyVVUQ to explore an input parameter space by generating ensembles of simulations and to collate the results into an easily analysable object. To distribute the large number of simulations to be run in each ensemble, we use FabSim3 to automate the submission of jobs on remote HPC resources. 

% {\color{blue}
% \small
% \begin{enumerate}
% 	\setlength\itemsep{-0.1em}
% \item A small Tube Map image specific to this application (ask Derek if you need a template).
% \end{enumerate}
% }
%\desc{The input will be provided by Robert Sinclair}

%-------------------------------------------
% 			UrbanAir
%-------------------------------------------
\subsection{Urban air pollution}

%Modelling air quality in complex urban can be a time consuming and tricky process. 
Predicting air quality in complex urban areas is a challenging field where researchers need to balance accuracy against a practical turnaround time. There are numerous models for predicting contamination 
transport and dispersion, ranging from simple Gaussian models (that are fast and cheap but not necessarily accurate) to computational fluid dynamics simulations that are slower, more costly and potentially very accurate. 

In all cases the most important, and often missing part, is an accurate emission database that
contains pollutant emission rates for different types of sources: point (e.g. industrial chimneys), line (e.g. road transportation) and areas 
(e.g. house heat appliances). The pollutants we consider are $NO_2$/$NO_x$, $SO_2$ and two types of particulate matter (also known as floating dust): $PM_{2.5}$ for 
particles 2.5$ \mu m$ or less in diameter and $PM_{10}$ for particles 10$ \mu m$ or less in diameter.
We use a mesoscale weather prediction model WRF~\cite{wrf_2011} and an all-scale geophysical flow solver EULAG~\cite{prusa2008eulag} in our multiscale
simulation~\cite{wrf_eulag_2012} to address  uncertainties coming from incomplete emission database, and to provide a quantitative air quality 
prediction~\cite{inpress_confidence}.

We use VECMAtk to automate sampling, ensemble generation, execution of simulation codes on HPC machines, collating results and post-execution analysis. This automation simplifies the provision of better prediction results, because we analyze many different initial conditions and can automatically provide mean results (or some weighted ones) from all simulations. We are also able to run sensitivity analysis of the input parameters to identify and select the most important ones, thus limiting the number of required ensembles for future runs.

\begin{figure}[ht]
    \centering
    \includegraphics[scale=0.60]
    {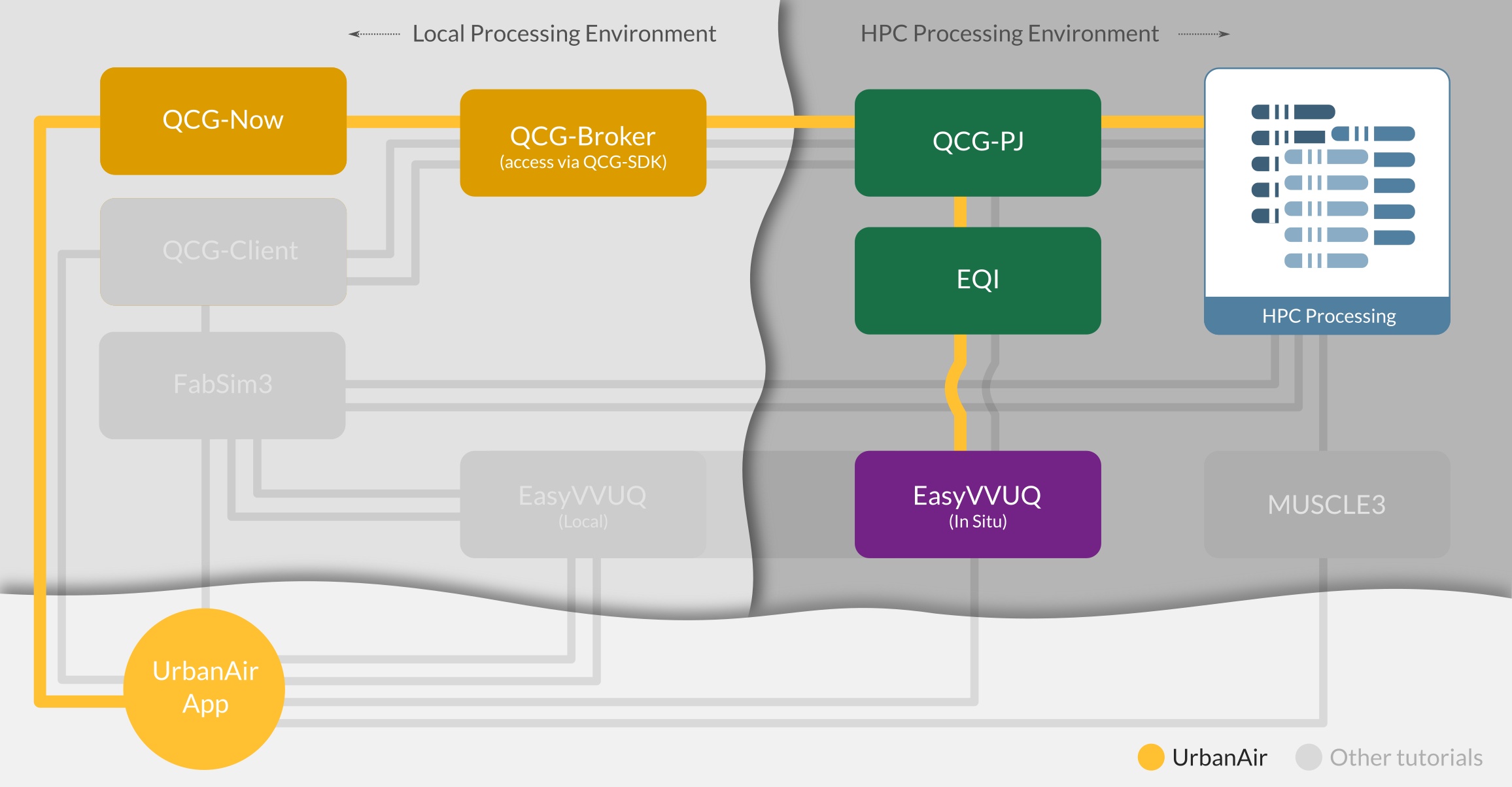}
	\caption{“Tube map” showing which VECMAtk components are used by the UrbanAir application.}	
	\label{UrbanAirMap}
	\medskip
\end{figure}

In terms of tools we rely on EasyVVUQ, coupled with QCG-PJ using EQI. Using Python we define the parameter space to be sampled, application to be run in HPC environment, hardware requirements (e.g. number of nodes and cores), and the parameters to be analysed after simulations ended. We then use QCG client and QCG-PJ to submit and control jobs on the HPC resources, EasyVVUQ to generate the samples, EQI for enabling the samples to be executed with QCG-PJ and EasyVVUQ to analyze the results. We summarize our workflow in Figure \ref{UrbanAirMap}.

%-------------------------------------------
% 			In-stent Restenosis
%-------------------------------------------
\subsection{Biomedicine}

In-stent restenosis (ISR) is the renewed occurrence of arterial stenosis (narrowing of an artery) after it was initially treated by installing a metal stent, as a result of excessive cell growth. The results of uncertainty and sensitivity analysis for a two-dimensional model of in-stent restenosis (ISR2D) are presented in \cite{Nikishova2018}. Improved effectiveness of uncertainty propagation was achieved by applying a semi-intrusive metamodelling method and the results are demonstrated in \cite{Nikishova_2019_Semi-intrusive} and in \cite{Ye_2020-intrusive}.

Current research on the design of the UQ and SA experiments for the three-dimensional model of the ISR model (ISR3D) is ongoing. The ISR3D model is implemented using MUSCLE~3 discussed in Section 4\ref{sec:MUSCLE3}, which allows performing more advanced semi-intrusive algorithms. In particular, we plan to train a metamodel on the fly and combine that with a cross-validation test of the effect of the approximation error on the results of the micro model.

%-------------------------------------------
% 			COVID-19
%-------------------------------------------
\subsection{Coronavirus modelling}

Since 2019, the world has been severely affected by the spread of the SARS-CoV-2 coronavirus, and the COVID-19 disease it causes. Although many models are able to approximate the viral spread on the national level, few solutions are available to forecast the spread on a local level, e.g. in towns or city boroughs. Within the HiDALGO project~\footnote{http://hidalgo-project.eu} we are working on the Flu And Coronavirus Simulator (FACS)~\footnote{https://facs.readthedocs.io} to help address this challenge.

We use VECMAtk primarily to systematically validate the code, analyse parameter sensitivities, execute ensemble simulations to account for aleatory uncertainty in the code, and to easily produce ensemble forecasts which involve a wide range of scenarios, applied to a range of boroughs in London. 

To achieve this, we primarily rely on FabCovid19, which is a FabSim3 plugin, as well as QCG-PJ. Both tools combined allow us to automate these tasks, and seamlessly create, run and post-process ensemble simulations. We also use EasyVVUQ to analyse parameter sensitivities, which we present in the next subsection.

%This workflow is demonstrated in Figure \ref{FabCovid19Map}.

%\begin{figure}[ht]
%    \centering
%    \includegraphics[trim=10 10 10 10,clip=true,scale=0.30]
%    {./Figs/FabCovid19Map.pdf}
%	\caption{“Tube map” showing which VECMAtk components are used by the COVID19 application.}	
%	\label{FabCovid19Map}
%	\medskip
%\end{figure}

%-------------------------------------------
% 			Demonstration of a single example VVUQ analysis
%-------------------------------------------
\subsection{Demonstration of a single example VVUQ analysis: COVID-19 application}

To demonstrate the added value offered by VECMAtk more concretely, we showcase one specific VVUQ procedure example, using the Flu And Coronavirus Simulator. We perform sensitivity analysis across six different input parameters of FACS to identify their sensitivity relative to our quantity of interest (QoI). In Table \ref{FACS_param_range} we provide the default value for each parameter along with the range of likely values. We use the Chaospy library\footnote{https://pypi.org/project/chaospy/} in EasyVVUQ to generate samples from the input parameters. Specifically, in this example we used the stochastic collocation method with a sparse-grid sampling plan of 15,121 samples, which we then convert to simulation inputs using the EasyVVUQ encoder, and submit to the HPC resource using FabSim3. Once execution has concluded, we then decode and collate the results and perform a Sobol sensitivity analysis relative to our QoI (number of deaths over time).

\begin{table}[H]
\centering

\begin{tabular}{lccc}
\hline
Parameters & Type &  Default value & Uniform range\\\hline
infection rate & float &  0.07 & (0.0035, 0.14)\\
mortality period & float &  8.0 & (4.0, 16.0)\\
recovery period & float &  8.0 & (4.0, 16.0)\\
mild recovery period & float &  8.05 & (4.5, 12.5)\\
incubation period & float &  3.0 & (2.0, 6.0)\\
period to hospitalisation & float &  12.0 & (8.0, 16.0)\\\hline
\end{tabular}
\caption{Defining parameter space for the uncertain parameters of the Flu And Coronavirus Simulations (FACS).}
\label{FACS_param_range}
\end{table}

\begin{figure}[ht]
    \centering
		 % 								trim = left bottom right up	 	    
    \includegraphics[trim=10 100 1 30,clip=true,scale=0.50]
    {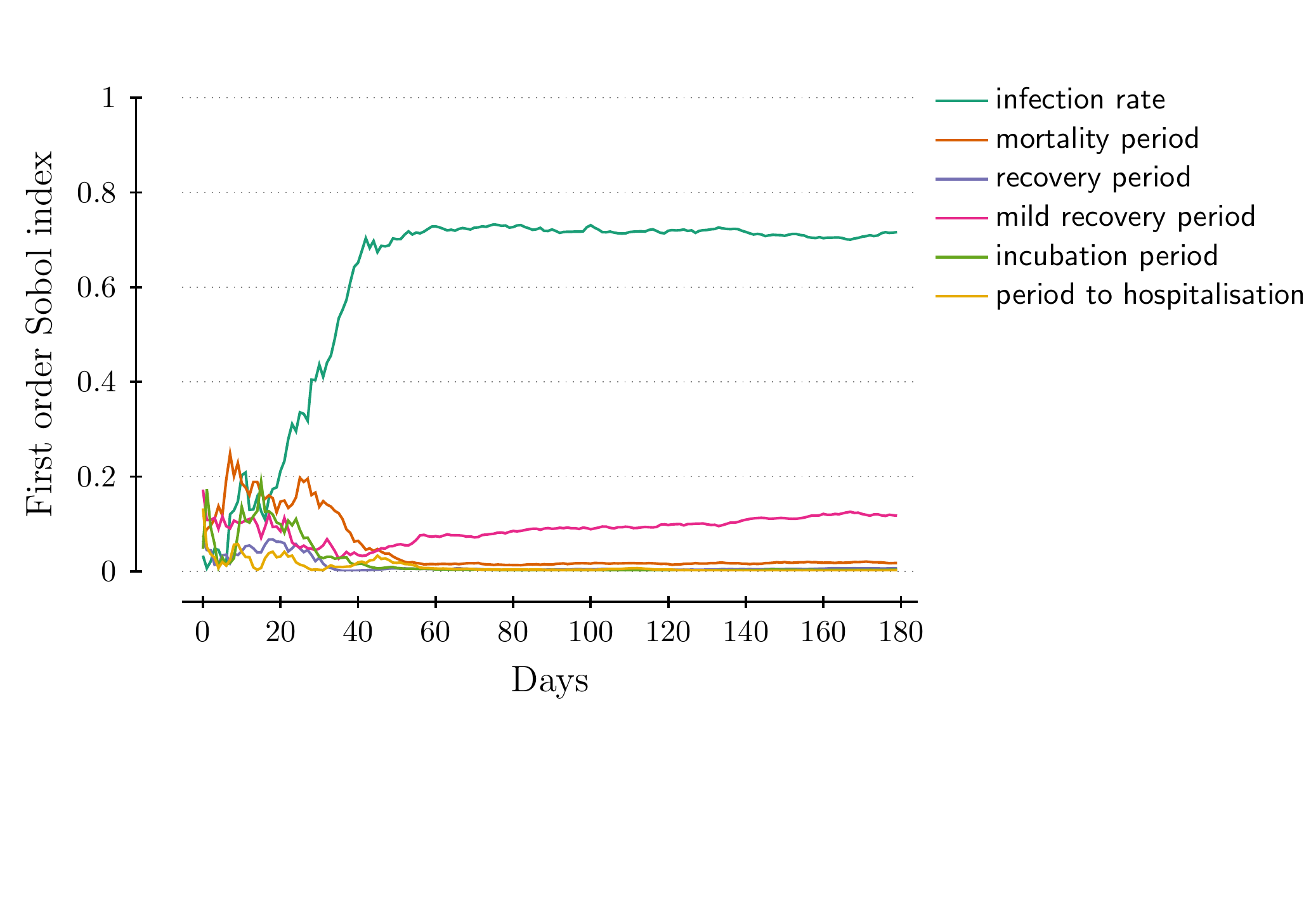}
	\caption{The first order Sobol indices for each of the uncertain parameters of the Flu And Coronavirus Simulations (FACS) for the London Borough of Brent.}	
	\label{sobols_first_dead}
	\medskip
\end{figure}

We present the first-order Sobol sensitivity indices for each parameter in Table \ref{FACS_param_range} and in Figure \ref{sobols_first_dead}. These global sensitivity indices \cite{saltelli2008global} measure the fraction of the output variance (over time here), that can be attributed to a single input parameter. During the first twenty days, all parameters except for the (non-mild) recovery period have a significant effect on the number of deaths. However, as the simulation progresses, the number of deaths is mainly sensitive to the infection rate (which describes how quickly the infection spreads) and the mild recovery period (which describes how long people with mild COVID remain ill and infectious).

%-------------------------------------------
% 			Toolkit performance and scalability
%-------------------------------------------
\section{Toolkit performance and scalability}\label{S:scalability}	

The VECMA toolkit has been developed for use with large supercomputers, and therefore needs to be both fast and scalable. In this section, we present a range of key performance aspects of the toolkit, and discuss the advantages and limitations of VECMAtk in terms of scalability and exascale readiness.

The main performance-critical aspects of the toolkit involve:
\begin{enumerate}[(a)] 
    \item The sampling of parameter values and creation of large numbers of simulation inputs.
    \item The submission, execution and retrieval of large ensembles of simulation jobs.
    \item The efficient movement of data between local and remote resources.
    \item The efficient movement of data between coupled models.
\end{enumerate}

We will now briefly discuss the performance-related characteristics of VECMAtk in relation to these four potential bottlenecks:

\textbf{Sampling parameter values and creating large numbers of simulation inputs.} There are multiple ways in which EasyVVUQ helps with the issues involved in sampling many and expensive simulation outputs. First of all, EasyVVUQ uses an SQL backend database that can handle 10000s of samples or more. It also tracks the status of job execution, allowing users to cope with hardware failures by restarting failed jobs without having to rerun successful ones. Furthermore, EasyVVUQ allows one to sample in stages - appending more samples if results are not yet sufficiently robust, without losing the results already in the database. Lastly, EasyVVUQ can delegate the ensemble job submission to other scalable components, such as Dask or QCG-PJ. These components then execute part or all of the simulations on HPC resources.

\textbf{Submission and execution management of large ensembles of simulation jobs.} Many scientific applications need to run large ensemble simulations (1000+ runs) to perform UQ and SA, which cannot be executed as individual jobs on most supercomputers due to scheduler constraints, and require a pilot job mechanism such as QCG-PJ. QCG-PJ has been shown to efficiently execute 10000 jobs with less than 10\% overhead, even if those jobs only last for one second each~\footnote{See VECMA Deliverable 5.2: https://www.vecma.eu/wp-content/uploads/2019/12/VECMA\_D5.2\_First-Report-Infrastructure\_PSNC\_20191208.pdf}. 
%Using FabSim3, QCG-PJ can be integrated in a larger automated workflow, where local processing is combined with remote (ensemble) job submission. Running ensembles in this way can be more convenient (often a one-liner command on the local machine suffices), but also introduces some performance overhead, as data will need to be transferred between the local and remote resource. 

\begin{figure}[ht]
	\centering
		 % 								trim = left bottom right up	 			
		\subfigure[ensemble size = 15121, QCG-PilotJob= True]{{\includegraphics[trim=10 10 10 40,clip=true,scale=0.33]{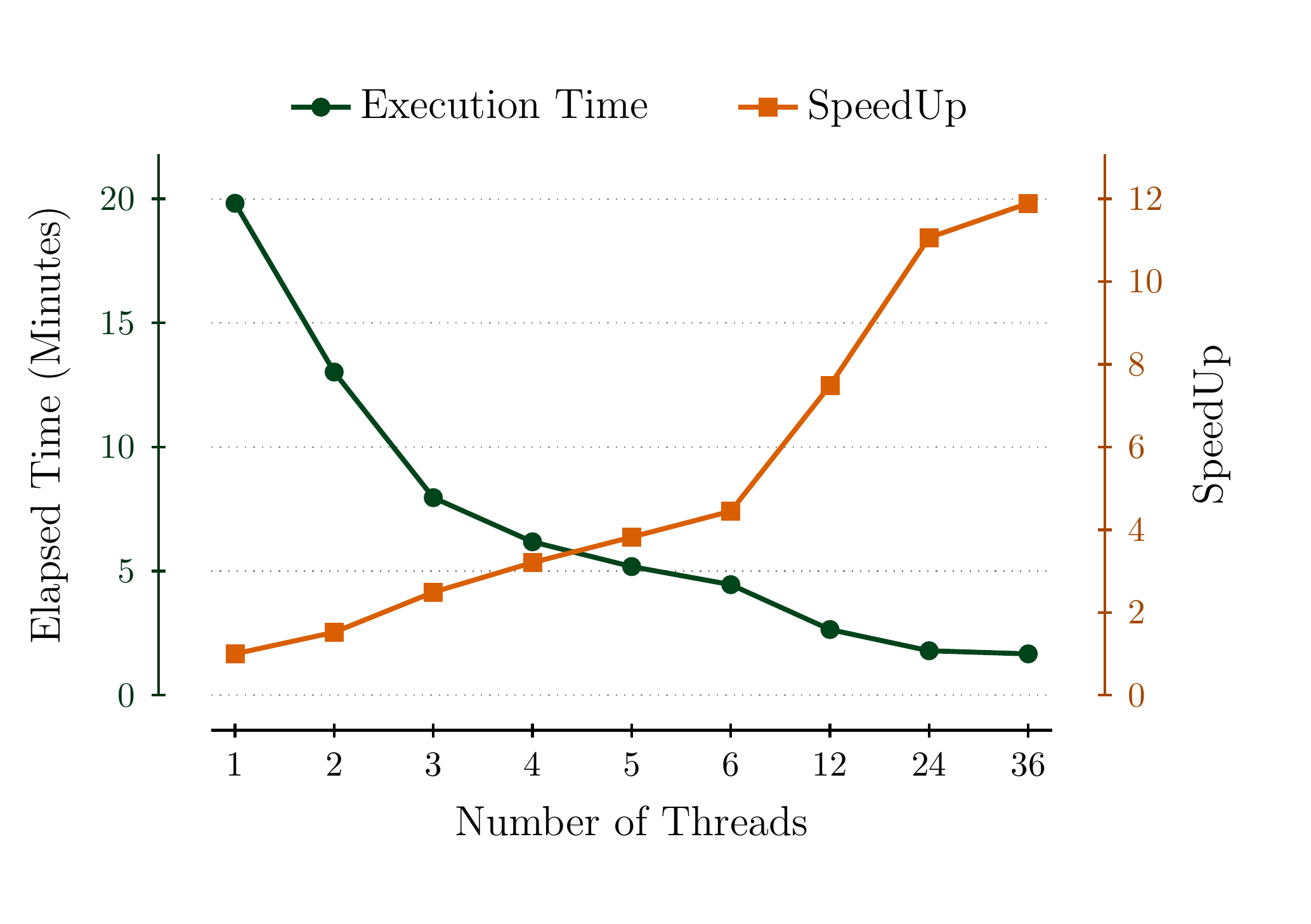}}}
		\subfigure[ensemble size = 4865, QCG-PilotJob= False]{{\includegraphics[trim=10 10 10 40,clip=true,scale=0.33]{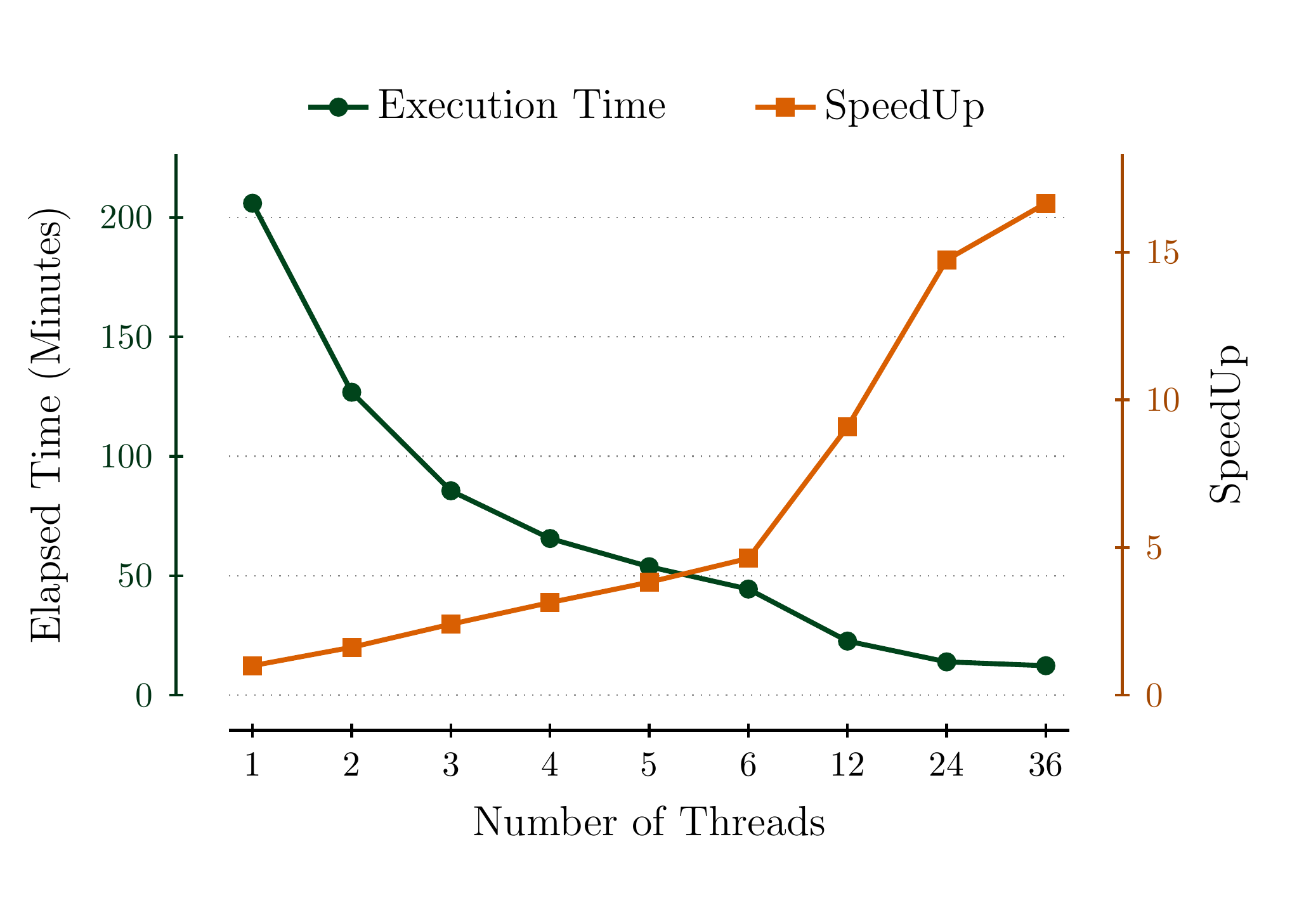}}}
	\caption{Time required to submit 15121/4865 jobs with FabSim3 (with/without QCG-PJ) relative to the number of job submission threads used. Graph is made using average of 10 repetition of each ensemble size. Please note that, here we only measure the job submission overhead, so, queuing time and job execution on computing nodes are not considered in our test.}
	\label{FabSim_scalability}
	\medskip % induce some separation between caption and explanatory material

	\end{figure}

To improve the FabSim3 ensemble submission performance, we integrated it with QCG-PJ and enabled multi-threaded job submission from the local machine. To demonstrate the benefit of this, we measured the total elapsed time of the job submission to a remote machine for the epidemiology \footnote{https://github.com/djgroen/FabCovid19} application. Due to limitation of maximum number of submited job per user \footnote{The maximum number of jobs a user can have running and pending at a given time is 5000. If this limit is reached, new submission requests will be denied until existing jobs in this association complete.} on the Eagle supercomputer, we use ensembles sizes of 4,865 and 15,121 runs, and use up to 36 threads for the submission process. We enabled pilot jobs for the larger ensemble, but not for the smaller ensemble. We present the results of this scalability test in Figure \ref{FabSim_scalability}.

For FabSim3 ensembles without the use of QCG-PJ, we find that the job submission overhead is 2-3 seconds per job when using a single thread, but decreases to less than a second per job when using four or more threads. When using FabSim3 with QCG-PJ, the job submission overhead reduces by a factor 2 for ensembles with 100 jobs to about 0.35 seconds per job. The overhead further diminishes for larger ensembles, as we measure an overhead of about 0.275 seconds per job for ensembles with 1000 jobs. In Table~\ref{Tab:pilotcompare} we compare the three main ensemble submission approaches in VECMA in terms of performance, usability and in regards to the need of any remote deployment.

\begin{table}
\begin{center}
 \begin{tabular}{c c c c c c} 
 \hline
 Approach & Usability & remote & remote  & \multicolumn{2}{c}{submission overhead}  \\
          &           & file stage? & deployment? & per 100 jobs & per 1000 jobs \\ 
 \hline
 FabSim3 only           & excellent & yes & Not needed      & 40-90s & not attempted\\
 FabSim3 + QCG-PJ & excellent & yes & user-space only & 33-36 s & 250-300 s\\
 QCG-PJ only      & good      & no & user-space only & $\lt$5 s & 40 s\\
 \hline
\end{tabular}
\caption{Comparison of the three main ensemble job submission approaches in VECMA, in terms of general usability, whether files are staged to and from the remote resource as part of the approach, the need for remote deployment work, and the overall performance.}
\label{Tab:pilotcompare}
\end{center}
\end{table}

\textbf{Efficient movement of data between local and remote resources.} Large simulations, as well as large ensembles of smaller simulations, are often accompanied by the production of large amounts of complex data. Within the VECMA project, we find that the organising and re-organising of this data is a cognitively intensive task that is prone to human error if not automated. Both the FabSim3 and EasyVVUQ tools provide a range of data structure conventions that transparently automate low-level data management aspects, allowing users to focus on the complexities that occur at higher levels. For instance, FabSim3 automatically curates input and output and while EasyVVUQ automates the encoding and decoding of simulation files. In addition, FabSim3 and QCG-Client automatically perform file staging to and from HPC resources.

In terms of performance, we chose to focus less on optimising file transfer rates (although FabSim3 and QCG-Client do support GridFTP), and more on limiting the number of file transfers altogether. The clearest example of this optimisation is the tight integration of EasyVVUQ directly with QCG-PJ. Using EQI, users can use HPC resources to generate, run and analyse their simulation ensembles for VVUQ purposes, and do this iteratively and dynamically within a single Python script. In this way, all the computations are brought to where the data is, and input files only need to be staged in once in advance, even if the workflow contains a large number of (dynamic) iterations.

\textbf{The efficient movement of data between coupled models.} MUSCLE 3 is primarily positioned to address this bottleneck, and provides added value for instance by eliminating the need for file I/O in the coupling.

%-------------------------------------------
% 			Conclusions
%-------------------------------------------
\section{Conclusions}\label{S:conclusions}

In this paper we have presented the VECMA toolkit for the verification, validation and uncertainty quantification (VVUQ) of single and multiscale applications. As showcased by the wide range of applications that use the toolkit already, VECMAtk is unique in its ability to combine a wide range of VVUQ procedures with a streamlined automation approach, and trivially accessible capabilities to execute large job ensembles on pre-exascale resources. 
In addition, VECMAtk components can be flexibly combined, allowing users to take advantage of parts of the toolkit while retaining a very limited deployment footprint.

As part of this paper, we have described a number of exemplar applications from a diverse range of scientific domains (fusion, forced migration, climate, advanced materials, urban air pollution, biomedical simulation and coronavirus modelling), 
all used by researchers today. All these examples are open source and accompanied with tutorials on http://www.vecma-toolkit.eu/, allowing new users to exploit them as building blocks for their own applications.

Through the VECMA toolkit, we aim to make VVUQ certification a standard practice and to make it simpler to quantify uncertainties, parameter sensitivities and errors in scientific simulations. VECMAtk is used to establish these procedures irrespective of the application domain, allowing key VVUQ algorithms to be reused across disciplines. Because VVUQ is essential to ensure that the key simulation results are indeed actionable, the VECMA toolkit is an important tool to help ensure that scientific simulations can be responsibly used to inform decision-making.

\enlargethispage{20pt}

\dataccess{}

\aucontribute{}

\competing{The authors declare that they have no competing interests.}

\funding{
This work was supported by the VECMA project, which has received funding from the European Union Horizon 2020 research and innovation programme under grant agreement No 800925. The development of MUSCLE3 and its respective description was supported by the Netherlands eScience Center and NWO under the e-MUSC project. The development of the migration and coronavirus modelling applications was supported by the EU-funded HiDALGO project (grant agreement No 824115). The calculations were performed in Poznan Supercomputing and Networking Center.
}

\ack{} 

\newpage
\appendix

%\section*{Appendix A. Changelog of VECMAtk components compared to the initial release.}

%\appendix

%%%%%%%%%% Insert bibliography here %%%%%%%%%%%%%%

\section*{Appendix A: improvements since first VECMAtk release}

In Table~\ref{tbl_vecmatk_imp} we present the fundamental improvements and changes in the development of each VECMAtk component since the first major release:

\begin{table}[ht]
\centering
\small
%\begin{tabular}{l|i{0.6\textwidth}}
\begin{tabular}{p{0.19\textwidth}|i{0.75\textwidth}}

    \toprule
    VECMAtk Component &  \multicolumn{1}{p{0.6\textwidth}}{List of improvements and changes in development} \\
    \midrule
    FabSim3   & 	\item Added support for multi-threaded job management
                	\item Added automated installation and configuration of FabSim3 on different OS
                	\item Fixed synchronizing/copying for a large number of files and folders
                	\item Vastly improved the performance of \texttt{run\_ensemble}
                	%\item Added support for replica execution (multiple simulations with identical input data and input parameters)
                	\item Revamped the documentation
    \\
    \midrule
    EasyVVUQ  &  \item Added support for vector-valued quantities of interest
                 \item Developed an extensive unit testing suite
                 \item Added support for Dask \footnote{https://dask.org} as an execution back-end
                 \item Improved sparse-grid Stochastic Collocation sampler
                 \item Dimension-adaptive Stochastic Collocation sampler 
    \\
    \midrule
    QCG Pilot Job &   \item Developed the node agent for more efficient launching of jobs across many nodes
                      \item Added more optimal way of handling iterative jobs
                      \item Increased the test base and test code coverage
                      \item Simplified execution of bash scripts as Pilot Job's tasks
                      \item Bugfixes for execution of multi-node Pilot Jobs and OMP jobs
    \\
    \midrule
    MUSCLE3 &  \item Newly added; contributed by the e-MUSC project
               \item Submodel coupling for spatial and/or temporal scale separation
               \item Submodel instance sets and ensembles
               \item Messages may contain grids, lists, dictionaries and basic types
               \item Support for Python, C++, Fortran and MPI, with tutorials and API documentation
               \item Automatic building of its dependencies on a variety of platforms
    \\
    \midrule
    EasyVVUQ-QCGPilotJob &  \item Provided a new API that simplifies the execution of EasyVVUQ scenarios using the QCG Pilot Job system.
                     \item Introduced a new mechanism to support custom encoders usage.
                     \item Made available a detailed tutorial that can be helpful for new users.
    \\
    \midrule
    QCG-Now &  \item  Developed an internal view for monitoring of applications (integration with QCG-Monitoring).
            \item Provided quick-template functionality for frequently submitted tasks. 
    \\
    \bottomrule
\end{tabular}
\caption{The list of modifications and enhancements in the development of each VECMAtk component since the first annual release in June 2019}
\label{tbl_vecmatk_imp}
\end{table}

\end{document}